# SOFT DISSIPATIVE MODES IN NEMATODYNAMICS

Arkady I. Leonov[a1], Valery S. Volkov[b]

[a] *Department of Polymer Engineering, The University of Akron, Akron, Ohio 44325-0301, USA.*

[b] *Laboratory of Rheology, Institute of Petrochemical Synthesis, Russian Academy of Sciences, Leninsky Pr., 29, Moscow 117912 Russia.*

**Abstract**

The paper analyses a possible occurrence of soft and semi-soft dissipative modes in uniaxially anisotropic nematic liquids as described by the five parametric Leslie-Ericksen-Parodi (LEP) constitutive equations (CE's). As in the similar elastic case, the soft dissipative modes theoretically cause no resistance to flow, nullifying the corresponding components of viscous part of stress tensor, and do not contribute in the dissipation. Similarly to the theories of nematic elastic solids, this effect is caused by a marginal thermodynamic stability. The analysis is trivialized in a specific rotating orthogonal coordinate system whose one axis is directed along the director. We demonstrate that depending on closeness of material parameters to the marginal stability conditions, LEP CE's describe the entire variety of soft, semi-soft and harder behaviours of nematic viscous liquids. When the only shearing dissipative modes are soft, the viscous part of stress tensor is symmetric, and LEP CE for stress, scaled with isotropic viscosity is reduced to a one-parametric, stress - strain rate anisotropic relation. When additionally the elongation dissipative mode is also soft, this scaled relation has no additional parameters and shows that the dissipation is always less than that in isotropic phase. Simple shearing and simple elongation flows illustrate these effects.

## 1. Introduction.

An important feature of liquids and solids with uniaxially symmetric rigid particles (molecules for molecular media or macroscopic particles for suspensions and nanocomposites) is that the particles have such an additional degree of freedom as their internal rotations. The problem on how to incorporate this effect in continuum approach was long a subject of many studies.

Many aspects of the problem were elaborated for liquids. A CE for isotropic fluids with spherical particles under action of internal angular momentum was first proposed by Born [1] and rediscovered later by Sorokin [2]. Grad [3] first analysed the non-equilibrium thermodynamic effects in such fluids. Then de Groot and Mazur

---

extensively discussed the problem in their text [4]. Ericksen [5] was first who developed the theory of viscous uniaxially anisotropic liquid with anisotropic yield caused by the orientation of the medium at the rest state. He used in his theory the condition of stress symmetry and, as additional state variable, director, i.e. unit vector that described internal rotations of uniaxially symmetric particles in liquid. In the following we will mention the model of Ericksen liquid as that without yield. One can also find an extensive review of other Ericksen papers regarding to the topic in the text by Truesdell and Noll [6] (Sect. 127). Allen and de Silva [7] extended the Ericksen approach. They analysed the macroscopic effects of rotating uniaxially symmetric particles including the internal rotational inertia of particles and internal couples, which caused non-symmetry of stress. Leslie [8] rationalized and simplified the theory [7] to what is known today as the Leslie-Ericksen model, where the stress tensor is generally non-symmetric. A detailed thermodynamic derivation [8] (see also de Gennes and Prost text [9], Ch.5) of CE's for flows of low molecular nematics, resulted in LEP CE's [8] with non-symmetric stress. The text [9] demonstrated for the nematic liquid crystals the contributions of the external magnetic or electric fields and effects of space gradient of the director in internal couples ("molecular fields").

De Gennes [10] analyzed first the behavior of liquid crystalline (LC) crosslinked elastomers as weekly elastic nematic solids, including into consideration the director gradient. In many publications after that Warner and coworkers (e.g. see the paper [11] and references there) developed a theory with finite strains, and finite internal rotations, omitting however, the space gradient director terms. Remarkably, this approach leads to a good comparison of theory with experiments for LC elastomers.

A unique phenomenon, unusual for conventional anisotropic elasticity, has been discovered and confirmed experimentally for nematic elastomers. It is the occurrence of the *soft deformation modes*, which theoretically cost no elastic energy when accompanied by director rotation. These soft deformations analyzed in papers [11-13] and others on the basis of a particular CE, present a special case of the soft mode class for anisotropic materials with internal rotations, predicted first by Golubovich and Lubensky [14] (1989).

The present paper analyses the conditions for occurrence of dissipative soft deformation modes in theories of viscous nematic liquids. The examples of these liquids include the cases of LC nematics and suspensions of uniaxially symmetric particles. The outline of the paper is as follows. We first briefly discuss, with a little new interpretation, the well-known equations for kinematics and dynamics of nematic liquids, as well as LEP CE's. We also briefly discuss the well-known transition from LEP to Ericksen CE's when the asymmetry in stress tensor is neglected. We then analyze the stability conditions and a special condition of marginal instability, with consequences for formulation of CE's. Finally, we briefly exemplify our general results for simple shear and simple elongation flows.

## 2. Kinematics and balance relations for internal rotations

All the continuum theories assume that it is possible to describe macroscopically the effects of orientation and internal rotations of particles as evolution of the unit vector $\underline{n}$ called director, and its rigid rotations [7] with angular velocity $\underline{\omega}^I$. It is convenient to decompose the angular velocity $\underline{\omega}^I$ in the sum of two vectors, the spin $\underline{\omega}^I_\parallel$ and an additional vector $\underline{\omega}^I_\perp$, i.e. $\underline{\omega}^I = \underline{\omega}^I_\parallel + \underline{\omega}^I_\perp$, the spin vector being directed along the director, i.e. $\underline{\omega}^I_\parallel = \omega^I_\parallel \underline{n}$. It characterises on average the angular velocity of revolution of particles around their axis. The (scalar) angular velocity of spin $\omega^I_\parallel$ is then defined as $\omega^I_\parallel = \underline{\omega}^I \cdot \underline{n}$. Then the additional component $\underline{\omega}^I_\perp$ characterises the change in the $\underline{\omega}^I$ caused by change in director orientation. Evidently, the vectors $\underline{\omega}^I_\parallel$ and $\underline{\omega}^I_\perp$ are orthogonal, i.e. $\underline{\omega}^I_\perp \cdot \underline{\omega}^I_\parallel = \omega^I_\parallel \underline{n} \cdot \underline{\omega}^I_\perp = 0$. The angular velocity of body (frame) rotations $\underline{\omega}$ can also be decomposed in the sum of orthogonal components, $\underline{\omega} = \underline{\omega}_\parallel + \underline{\omega}_\perp$, where $\underline{\omega}_\parallel = \omega_\parallel \underline{n} = (\underline{\omega} \cdot \underline{n})\underline{n}$.

Since the director exercises the rigid rotation, its speed $\underline{\dot{n}}$ is expressed as:

$$\underline{\dot{n}} = \underline{\omega}^I \times \underline{n} = \underline{\omega}^I_\perp \times \underline{n} = -\underline{\underline{\omega}}^I_\perp \cdot \underline{n}, \text{ or } \dot{n}_i = \varepsilon_{ijk}\omega^I_{\perp j}n_k = -\omega^I_{\perp ik}n_k. \qquad (1)$$

Here $\varepsilon_{ijk}$ is the antisymmetric unit tensor and hereafter overdot denotes the time derivative. It is also convenient to introduce the anti-symmetric tensors, $\underline{\underline{\omega}}^I, \underline{\underline{\omega}}^I_\parallel$ and $\underline{\underline{\omega}}^I_\perp$, called respectively the total, spin and orientation internal 'vorticities', and also the similar anti-symmetric tensors $\underline{\underline{\omega}}, \underline{\underline{\omega}}_\parallel$ and $\underline{\underline{\omega}}_\perp$. Here the evident equalities hold: $\underline{\underline{\omega}}^I = \underline{\underline{\omega}}^I_\| + \underline{\underline{\omega}}^I_\perp$, and $\underline{\underline{\omega}} = \underline{\underline{\omega}}_\| + \underline{\underline{\omega}}_\perp$. The components of these tensors are defined as follows: $\omega^I_{ik} \equiv \varepsilon_{ikj}\omega^I_j$, $\omega_{ik} \equiv \varepsilon_{ikj}\omega_j$, and the same for their "$\perp$" and "$\|$" components. Here the body vorticity $\underline{\underline{\omega}}$ is commonly defined as the anti-symmetric part of the velocity gradient $\underline{\nabla}\underline{v}$ tensor whose symmetric part $\underline{\underline{e}}$ is the strain rate. According to the definitions of respective orthogonal vector components for total and internal angular velocities,

$$\underline{\underline{\omega}}^I_\| \cdot \underline{n} = \underline{\underline{\omega}}_\| \cdot \underline{n} = 0, \text{ therefore } \underline{\underline{\omega}}^I \cdot \underline{n} = \underline{\underline{\omega}}^I_\perp \cdot \underline{n}, \quad \underline{\underline{\omega}} \cdot \underline{n} = \underline{\underline{\omega}}_\perp \cdot \underline{n}. \tag{2}$$

The relative vorticity $\underline{\underline{\omega}}^r$ in the uniaxially symmetric anisotropic media is defined as:

$$\underline{\underline{\omega}}^r \equiv \underline{\underline{\omega}} - \underline{\underline{\omega}}^I = \underline{\underline{\omega}}^r_\| + \underline{\underline{\omega}}^r_\perp. \tag{3}$$

Note that the relative vorticity $\underline{\underline{\omega}}^r$ is frame invariant. Multiplying eq.(3) scalarly by $\underline{n}$ and using (1) and (2) yields:

$$\underline{\underline{\omega}}^r \cdot \underline{n} = \underline{\underline{\omega}}^r_\perp \cdot \underline{n} = \overset{0}{\underline{n}}, \quad \overset{0}{\underline{n}} \equiv \underline{\dot{n}} - \underline{n} \cdot \underline{\underline{\omega}}. \tag{4}$$

Here $\overset{0}{\underline{n}}$ is the co-rotational (Jaumann) tensor time derivative of director $\underline{n}$. Eqs.(1) and (2) also allow expressing $\underline{\underline{\omega}}^I_\perp$ and $\underline{\underline{\omega}}^I_\perp$ via $\underline{n}$ and $\underline{\dot{n}}$ as:

$$\omega^I_{\perp i} = \varepsilon_{ijk} n_j \dot{n}_k, \quad \underline{\underline{\omega}}^I_\perp = \underline{n}\underline{\dot{n}} - \underline{\dot{n}}\underline{n} \quad (|\underline{\underline{\omega}}^I_\perp| = |\underline{\dot{n}}|). \tag{5a}$$

Similar expressions for the $\underline{\underline{\omega}}^r_\perp$ and $\underline{\underline{\omega}}^r_\perp$ via $\underline{n}$ and $\overset{0}{\underline{n}}$ following from (2) and (4) are:

$$\omega^r_{\perp i} = \varepsilon_{ijk} \overset{0}{n}_j n_k, \quad \underline{\underline{\omega}}^r_\perp = \underline{n}\overset{0}{\underline{n}} - \overset{0}{\underline{n}}\underline{n}. \tag{5}$$

Formulae (5) are independent of physical models of nematic media.

The general macroscopic balance relation for the media with the uniaxial symmetry has been considered in paper [15]. In the case of LEP theory, the macroscopic relation for angular momentum can be presented in the simplified form:

$$\rho\underline{\underline{\dot{S}}} = 2\underline{\underline{\tilde{\sigma}}}^a; \quad \underline{\underline{\tilde{\sigma}}}^a \equiv \underline{\underline{\sigma}}^a + 1/2\rho\underline{\underline{m}} \quad S_{ij} = \varepsilon_{ijk}L_k. \tag{6}$$

Here $\underline{\underline{S}}$ is the internal moment of momentum per mass unit, $\underline{L}$ is the internal angular moment per mass unit, which represents averaged microscopic angular moment for particles, $\underline{\underline{m}}$ is the internal torque per mass unit, $\rho$ is the density, and $\underline{\underline{\sigma}}^a$ is the anti-symmetric part of the "irreversible" stress tensor obtained (as in the case of liquid crystals) by extracting the equilibrium stress from the full stress.

In isotropic molecular media, the density of internal torques $\underline{m}$, the key issue in the Cosserat non-symmetrical isotropic elasticity [6] (Sect.98), can be omitted in eq.(6) due to their negligible contribution in the balance equations and in the entropy production [3]. Perhaps because of the results [3], the non-symmetry effects of this theory for molecular isotropic liquids and solids are very small. However, in the nematic case, the internal torques, related to the molecular field, play important role in the theory of low molecular nematic liquid crystals [9].

In the macroscopic continuum theory, the internal angular momentum is represented as follows:

$$\underline{L} = \underline{\underline{I}} \cdot \underline{\omega}^I, \quad \underline{\underline{I}} = I_\perp \underline{\underline{\delta}} + (I_\| - I_\perp)\underline{n}\,\underline{n}. \tag{7a}$$

Here $I_\perp$ and $I_\|$ are averaged parameters of inertia moment per mass unit for the uniaxially symmetric particles. Substituting the decomposition $\underline{\omega}^I$ into (7) and using the second equation in (9) yields:

$$\underline{L} = I_\perp \underline{\omega}_\perp^I + I_\| \underline{\omega}_\|^I, \quad \underline{\underline{S}} = I_\perp \underline{\underline{\omega}}_\perp^I + I_\| \underline{\underline{\omega}}_\|^I. \tag{7}$$

Direct calculations show that

$$\underline{\dot{L}} \cdot \underline{n} = I_\| \dot{\omega}_\|^I, \quad \underline{\underline{\dot{S}}} \cdot \underline{n} = -I_\perp [\underline{\ddot{n}} + \underline{n}|\underline{\dot{n}}|^2] + I_\| \omega_\|^I \underline{\omega}_\perp^I. \tag{8}$$

The second equation in (8) demonstrates that the vectors $\underline{\underline{\dot{S}}} \cdot \underline{n}$ and $\underline{n}$ are orthogonal. Substituting (8) into (6) yields:

$$(\rho/2)[-I_\perp(\underline{\ddot{n}} + \underline{n}|\underline{\dot{n}}|^2) + I_\| \omega_\|^I \underline{\omega}_\perp^I] = \underline{\underline{\tilde{\sigma}}}^a \cdot \underline{n}, \quad \rho I_\| \dot{\omega}_\|^I = \varepsilon_{ijk} n_i \tilde{\sigma}_{jk}^a. \tag{9}$$

Kinetic energy per mass unit is the sum of common translation and internal rotation components. According to (6), the rotational component $K_r$ of kinetic energy (per mass unit) in the anisotropic system is:

$$2K_r = I_\perp (\omega_\perp^I)^2 + I_\| (\omega_\|^I)^2 = tr[I_\perp (\underline{\underline{\omega}}_\perp^I)^2 + I_\| (\underline{\underline{\omega}}_\|^I)^2]. \tag{10}$$

It is easy to show that

$$\dot{K}_r = tr(I_\perp \underline{\underline{\omega}}_\perp^I \cdot \underline{\underline{\dot{\omega}}}_\perp^I + I_\| \underline{\underline{\omega}}_\|^I \cdot \underline{\underline{\dot{\omega}}}_\|^I) = tr(\underline{\underline{S}} \cdot \underline{\underline{\omega}}^I).$$

Then the balance of kinetic energy of internal rotations is represented in the common form [4]:

$$(\rho/2)\dot{\underline{\underline{K}}}_r = (\rho/2)tr(\underline{\underline{S}} \cdot \underline{\underline{\omega}}^I) = tr(\underline{\underline{\tilde{\sigma}}}^a \cdot \underline{\underline{\omega}}^I). \tag{11}$$

## 3. Dissipation and anisotropic viscous constitutive equations

We now consider the incompressible case ($tr\underline{\underline{e}} = 0$) of nematic liquids with the density (per mass unit) of the Helmholtz free energy being of the form:

$$f = f(T, \underline{n}, \nabla \underline{n}). \tag{12}$$

Eq.(12) is valid for low molecular nematics. When $f$ is quadratic in $\nabla \underline{n}$, eq.(12) describes the well-known case of Frank elasticity. Assuming that the non-equilibrium (irreversible) part $\underline{\underline{\sigma}}$ of the total extra stress is viscous, the total extra tensor $\underline{\underline{\sigma}}_t$ can be split into symmetric and antisymmetric parts:

$$\underline{\underline{\sigma}} = \underline{\underline{\sigma}}^s + \underline{\underline{\sigma}}^a; \quad \underline{\underline{\sigma}}^s = \underline{\underline{\sigma}}^s_t - \underline{\underline{\sigma}}^s_{eq}; \quad \underline{\underline{\sigma}}^a = \underline{\underline{\sigma}}^a_t - \underline{\underline{\sigma}}^a_{eq}. \tag{13}$$

Here $\underline{\underline{\sigma}}^s_t$ and $\underline{\underline{\sigma}}^s_{eq}$ are the symmetric parts of the total extra stress tensor and equilibrium stress tensor, $\underline{\underline{\sigma}}^a_t$ and $\underline{\underline{\sigma}}^a_{eq}$ the same for anti-symmetric parts. Thus in case of molecular nematics with the free energy function (12), the internal (vector) torque $\underline{m}$ is represented through the "molecular (vector) field" $\underline{h}$ as [9]: $\rho \underline{m} = \underline{n} \times \underline{h}$ (or $\rho \underline{\underline{m}} = \underline{n}\underline{h} - \underline{h}\underline{n}$). Therefore the term $\underline{\underline{\tilde{\sigma}}}_a$ in eqs.(6), (9) and (11) is represented as:

$$\underline{\underline{\tilde{\sigma}}}^a = \underline{\underline{\sigma}}^a - 1/2(\underline{h}\underline{n} - \underline{n}\underline{h}) \tag{14}$$

Here $\underline{h}$ is calculated using the free energy and includes the magnetic field effect.

It should be noted that the detailed derivation [9] of dissipation function for low molecular weight nematic missed the contribution of internal rotation in the kinetic energy, $K_r$. Therefore the results [9] could be valid only when dynamic effects of internal rotations are negligible as compared to the static and external field effects in LC nematics. It formally means that neglecting the dynamics of internal rotations in the liquid crystal case, results in the equality

$$\underline{\underline{\tilde{\sigma}}}^a = 0, \text{ or } \underline{\underline{\sigma}}^a = 1/2(\underline{h}\underline{n} - \underline{n}\underline{h}). \tag{14a}$$

Namely eq.(14a) has been used for the derivation of LEP CE's (e.g. see [9]).

Along with liquid crystals, the second case with a particular version of free energy (12), $f = f(T, \underline{n}) = f(T)$, may be used for continuum description of viscous liquids with suspended uniaxially symmetric particles. Here the equilibrium molecular terms are absent, i.e. $\underline{\underline{\sigma}}^a_{eq} = 0;\ \underline{\underline{\sigma}}^s_{eq} = 0;$ and in the absence of external fields, $\underline{h} = 0$. Therefore in this case, $\underline{\underline{\tilde{\sigma}}}^s = \underline{\underline{\sigma}}^s$ and $\underline{\underline{\tilde{\sigma}}}^a = \underline{\underline{\sigma}}^a$. Dynamic effects in these suspensions have been considered in paper [16].

Using (12) the dissipation $D$ in the system is represented as follows:
$$D \equiv TP_S\big|_T = tr(\underline{\underline{\sigma}}^s \cdot \underline{\underline{e}}) - tr(\underline{\underline{\sigma}}^a \cdot \underline{\underline{\omega}}^r) \quad (\geq 0), \tag{15}$$

Here $P_S$ is the entropy production, $\underline{\underline{\omega}}^r$ is the relative vorticity defined in eq.(5), and the quantities $\underline{\underline{\sigma}}^s$ and $\underline{\underline{\sigma}}^a$ defined in (13) represent the non-equilibrium (viscous) parts of the extra stress tensor.

We now consider the CE's that follow from eq.(15). Using the quasi-linear approach of non-equilibrium thermodynamics [4] where the kinetic tensors depend on temperature and director, yields the general quasi-linear CE's:

$$\underline{\underline{\sigma}}^s = a_0 \underline{\underline{e}} + a_1(\underline{nn} \cdot \underline{\underline{e}} + \underline{\underline{e}} \cdot \underline{nn}) + 2a_2 \underline{nn}(\underline{nn} : \underline{\underline{e}}) + a_3(\underline{nn} \cdot \underline{\underline{\omega}}^r - \underline{\underline{\omega}}^r \cdot \underline{nn}) \tag{16a}$$

$$\underline{\underline{\sigma}}^a = a_3(\underline{nn} \cdot \underline{\underline{e}} - \underline{\underline{e}} \cdot \underline{nn}) + a_4(\underline{nn} \cdot \underline{\underline{\omega}}^r + \underline{\underline{\omega}}^r \cdot \underline{nn}). \tag{16b}$$

The form of terms in eqs.(16) is dictated by the symmetry of $\underline{\underline{\sigma}}^s$, anti-symmetry of $\underline{\underline{\sigma}}^a$, $\underline{n} \to -\underline{n}$ invariance, and the Onsager symmetry of kinetic coefficients. The scalar parameters $a_k$ in (16) are temperature dependent, i.e. $a_k = a_k(T)$. The possible Born's [1] term $a_5 \underline{\underline{\omega}}^r$ to the anti-symmetric part of the viscous stress might also occur in (16b). This term is of non-nematic nature and is presented even in isotropic state. The effect of this term has been analyzed in [15] and is responsible for the spin rotations in nematic liquids. Since in isotropic state its contribution in stress is negligible, we assume that this also happen in anisotropic state and neglect this term in the following.

Using eqs.(4) and (5), represents (16) in the identical form:

$$\underline{\underline{\sigma}}^s = a_0 \underline{\underline{e}} + a_1(\underline{nn} \cdot \underline{\underline{e}} + \underline{\underline{e}} \cdot \underline{nn}) + 2a_2 \underline{nn}(\underline{nn} : \underline{\underline{e}}) - a_3(\underline{n}\overset{0}{\underline{n}} + \overset{0}{\underline{n}}\underline{n}) \tag{17a}$$

$$\underline{\underline{\sigma}}^a = a_3(\underline{nn} \cdot \underline{\underline{e}} - \underline{\underline{e}} \cdot \underline{nn}) + a_4(\underline{n}\overset{0}{\underline{n}} - \overset{0}{\underline{n}}\underline{n}). \tag{17b}$$

Eqs. (17) are the same as used in incompressible case by de Gennes and Frost [9] (see eqs.5.27 and 5.28 there). The relations between the Leslie-Ericksen parameters $\alpha_k$ and the parameters $a_k$ is:

$$\alpha_1 = 2a_2, \; \alpha_2 = -a_3 - a_4, \; \alpha_3 = -a_3 + a_4, \; \alpha_4 = a_0, \; \alpha_5 = a_1 + a_3, \; \alpha_6 = a_1 - a_3.$$

Here additional Parodi equality, $\alpha_2 + \alpha_3 = \alpha_6 - \alpha_5 \; (= -2a_3)$, is satisfied identically. It should be mentioned that Parodi [16] derived first the equations (16), (17) in terms of the coefficients $\alpha_k$.

Substituting (17) into (9) with the use of (14) yields:

$$\rho[I_\perp(\underline{\ddot{n}} + \underline{n}|\underline{\dot{n}}|^2) - I_\parallel \omega_\parallel^I \underline{\omega}_\perp^I] = 2a_3[\underline{\underline{e}} \cdot \underline{n} - \underline{n}(\underline{\underline{nn}} : \underline{\underline{e}})] - 2a_4 \overset{0}{\underline{n}} - \underline{h}^\perp; \; \underline{h}^\perp \equiv \underline{h} - \underline{n}(\underline{h} \cdot \underline{n}) \quad (18)$$

Here $\underline{\omega}$ is the angular velocity vector of the body (frame). Additionally to (18) there is also equation, $(\rho/2)I_\parallel \dot{\omega}_\parallel^I = 0$, that shows that in the absence of the Born term $a_5 \underline{\omega}^r$, the spin of internal rotations is absent. Eqs.(17) and (18) represent the closed set of viscous nematodynamic equations.

When one can neglect the inertia terms of internal rotation shown in the left-hand side of (18), the right-hand side of this equations are simplified to

$$1/2 \underline{h}^\perp = -a_4 \overset{0}{\underline{n}} + a_3[\underline{\underline{e}} \cdot \underline{n} - \underline{n}(\underline{\underline{nn}} : \underline{\underline{e}})], \quad (19)$$

eq.(17a) still holds, and eq.(17b) reduces to the first relation in (19), which coincides with eq.(5.32) in the text [9].

Substituting (16) into (15) yields:

$$D = a_0 |\underline{\underline{e}}|^2 + 2a_1(\underline{\underline{nn}} : \underline{\underline{e}}^2) + 2a_2(\underline{\underline{nn}} : \underline{\underline{e}})^2 - 4a_3[\underline{\underline{nn}} : (\underline{\underline{e}} \cdot \underline{\underline{\omega}}^r)] - 2a_4[\underline{\underline{nn}} : (\underline{\underline{\omega}}^r)^2]. \quad (20)$$

Here the dissipation is expressed as quadratic form relative to the variables $\underline{\underline{e}}$ and $\underline{\underline{\omega}}^r$.

## 4. Stability conditions and dissipative deformational modes

The material parameters $a_k$ in (20) present the scaling factors for independent basis tensor invariants. Therefore the following conditions,

$$a_k \neq 0 \; (k = 0, 1, 2, 3, 4) \quad (21)$$

are used below to avoid degenerations of CE's (17).

The dissipation function (20) should be positive in non-equilibrium (i.e. for flows of the liquid) and vanish in the equilibrium. This imposes some constraints (or stability conditions) on material parameters in (20). To reveal these conditions and make easier further analysis, we introduce a special, time dependent, rotating orthogonal coordinate system $\{\hat{\underline{x}}\}$, whose one axis, say $\hat{x}_1$, is directed along the axis of director, $\underline{n}$. In the coordinate system $\{\hat{x}_i\}$:

$$\underline{n} = \{1,0,0\}; \quad \underline{\underline{nn}} = \begin{pmatrix} 1 & 0 & 0 \\ 0 & 0 & 0 \\ 0 & 0 & 0 \end{pmatrix}; \quad \underline{\underline{e}} = \hat{\underline{\underline{e}}}; \quad \underline{\underline{\omega}}^r = \hat{\underline{\underline{\omega}}}^r. \tag{22}$$

The last relation in (22) uses the fact of frame invariance for the anti-symmetric tensor $\underline{\underline{\omega}}^r$, characterizing relative rotations. Then in $\{\hat{\underline{x}}\}$ coordinates, equation (20) for dissipation is represented in the form:

$$\hat{D} = (a_0 + 2a_1 + 2a_2)\hat{e}_{11}^2 + a_0(e_{22}^2 + \hat{e}_{33}^2 + 2e_{23}^2) + 2\sum_{k=2,3}[(a_0 + a_1)\hat{e}_{1k}^2 + 2a_3\omega_{1k}^r \hat{e}_{1k} + a_4(\omega_{1k}^r)^2] \quad (\geq 0)$$
$$. \tag{23a}$$

Hereafter all the tensor components in $\{\hat{\underline{x}}\}$, are marked by upper caps. Because of the incompressibility condition, not all the terms in (23a) are independent. Using this condition to express, for example, $\hat{e}_{33} = -(e_{11} + e_{22})$ and substituting it back into (23a), yields the equivalent formula for the dissipation,

$$\hat{D} = (3/2 a_0 + 2a_1 + 2a_2)\hat{e}_{11}^2 + 2a_0[(\hat{e}_{22} + 1/2 e_{11})^2 + e_{23}^2] \tag{23b}$$
$$+ 2\sum_{k=2,3}[(a_0 + a_1)\hat{e}_{1k}^2 + 2a_3\omega_{1k}\hat{e}_{1k} + a_4\omega_{1k}^2],$$

which now contains only independent terms.

CE's (17a,b) written in $\{\hat{\underline{x}}\}$ take the form:

$$\hat{\sigma}_{11} = (a_0 + 2a_1 + 2a_2)e_{11}; \quad \hat{\sigma}_{22} = a_0 e_{22}; \quad \hat{\sigma}_{33} = a_0 e_{33}; \quad \hat{\sigma}_{23} = a_0 e_{23} \tag{24a}$$

$$\begin{cases} \hat{\sigma}_{1k} = (a_0 + a_1)e_{1k} + a_3\omega_{1k}^r \\ \hat{\sigma}_{1k} = a_3 e_{1k} + a_4\omega_{1k}^r \end{cases} \quad (k = 2,3) \tag{24b}$$

Hereafter CE's for anti-symmetric stress components are written only for $k < j$.

We call each of nine pairs $\{\hat{e}_{ij}, \omega_{ij}\}$ of kinematic variables in (24) the $\{i,j\}$ deformational/rotational dissipative mode or simply $\{i,j\}$ dissipative mode. The dissipative modes in (2.11) with $i = j$ and $i \neq j$ are naturally called *normal* and *shearing*, respectively. In the following, we are mostly interested in the shearing dissipative modes $\{1,k\}/\{k,1\}$, which for simplicity will be denoted as $\{1,k\}$, and the special normal, *elongation* dissipative mode $\{11-22\}$ corresponding to the strain rate difference $e_{11} - e_{22}$.

The thermodynamic stability constraints are established when demanding the quadratic form in (23b) to be positive definite. Since all terms in (23b), including two last quadratic forms for $k=2$, $3$ $k=3$, are independent, the *necessary and sufficient* stability conditions are:

$$a_0 > 0; \quad a_4 > 0; \quad a_0 + a_1 > 0; \quad 3/4 a_0 + a_1 + a_2 > 0; \quad (a_0 + a_1)a_4 > a_3^2. \tag{25}$$

As follows from (21) and (25), the parameters $a_1, a_2$, and $a_3$ are sign indefinite. It should be noted that instead the third inequality in (25) Parodi [16] (see also [9]) obtained only *sufficient* stability condition, which in our notations is: $1/2 a_0 + a_1 + a_2 > 0$. This inequality clearly follows from (23a) under incorrect assumption that all the terms there are independent. As shown below, the necessary and sufficient stability conditions play a pivotal role for revealing existence of soft dissipative modes.

When the set of parameters $P = \{a_0, ..., a_4\}$ satisfies (2.7) and (2.12), there are always 1-1 dependences between the tensors $\underline{e}$, $\underline{\underline{\omega}}$ and $\underline{\underline{\sigma}}$ described by CE's (24). This fact is evident for CE's (24a). It is also clear for CE's (24b) if they are considered as a coupled set of two linear algebraic equations with positive determinant,

$$Det \equiv (a_0 + a_1)a_4 - a_3^2 > 0. \tag{26}$$

This result, proved for particular case (24) in $\{\hat{x}\}$, is clearly valid for the general (tensor) form of CE's (16).

## 5. Marginal stability and soft/semi-soft dissipative modes

The marginal stability corresponds to the physically possible cases when some inequalities in (25) are changed for equalities. Due to (21), only two marginal stability constraints can exist,

$$3/4 a_0 + a_1 + a_2 = 0, \tag{27}$$

$$Det \equiv (a_0 + a_1)a_4 - a_3^2 = 0. \tag{28}$$

We call the $\{i, j\}$ dissipative modes whose parameters satisfy (27), (28) or both, *marginally stable*.

It is also possible to consider the case when stability conditions (25) are satisfied, but two last inequalities in (25) are close to those in (27) and (28):

$$3/2a_0 + 2a_1 + 2a_2 = a_0 O(\delta), \tag{27a}$$

$$Det \equiv (a_0 + a_1)a_4 - a_3^2 = a_3^2 O(\delta). \tag{28a}$$

Here $0 < \delta \ll 1$. We call the $\{i, j\}$ stable nematic modes whose parameters $P$ satisfy (27a), (28a) or both, *nearly marginally stable*.

To compare the orders of stresses in possible (nearly) marginally stable modes we can generally accept that $|\underline{\underline{e}}| \sim |\underline{\underline{\omega}}^r| \sim e$. Here $e$ is a scalar specific strain rate characterising the intensity of flow. A particular $\{i, j\}$ dissipative mode is called *soft* if the viscous component of stress tensor $\hat{\sigma}_{ij} = 0$ while $\vec{e}_{ij}, \omega_{ij}^r = O_{ij}(e)$. CE's (24) show that the soft modes, if exist, are either the shearing $\{1,k\}$ described by CE's (24b), or the elongation $\{11\text{-}22\}$ one.

Consider first the possible dissipative soft shearing modes $\{1,k\}$, when $\hat{\sigma}_{1k} = \sigma_{k1} = 0$. Due to (24b) there are 1-1 dependences between relative rotations and respective strains in the shearing soft modes $\{1,k\}$, although these strains are not unique. Consider now the simple stretching along the axis $\hat{x}_1$. In this case the strain rate tensor $\underline{\underline{e}}$ is represented as: $\underline{\underline{e}} = \dot{\varepsilon} \cdot \underline{\underline{diag}}\{1, -1/2, -1/2\}$ where $\dot{\varepsilon}$ is the elongation strain rate. Then the elongation stress $\hat{\sigma}_{el} = \sigma_{11} - \sigma_{22}$, is calculated due to (24a) as:

$$\hat{\sigma}_{el} = (3/2a_0 + 2a_1 + 2a_2)\dot{\varepsilon} \tag{29}$$

Therefore if the elongation mode is soft, i.e. $\hat{\sigma}_{el} = 0$ when $\dot{\varepsilon} \neq 0$, equation (29) results in the marginal stability condition (27) and *vice versa*. Evidently, these possible soft dissipative modes do not contribute in the dissipation, i.e. if they exist they deliver a minimum of the dissipation in the parametric "space".

The *semi-soft* modes are defined as stable modes $\{i, j\}$ in CE (24), which for small positive numerical parameter $\delta$ satisfy the relations

$$\hat{\sigma}_{el}/G_0 = O(\delta) \text{ and } \hat{\sigma}_{1k}/G_3 = O(\delta), \text{ while } \vec{e}_{ij}, \omega_{ij} = O_{ij}(e). \tag{30}$$

The stresses for semi-soft modes, as well as their contribution in dissipation, are considerably smaller than for other modes.

The general behavior of LC nematics can now be formally classified in terms of values of parameter $\delta$, as soft ($\delta = 0$), semi-soft ($0 < \delta \ll 1$), and hard ($\delta > \sim 1$).

We now establish relations between the existence of soft (or semi-soft) modes and equations (27), (28) (or (27a), (28a)). Relation (29) clearly shows that the *dissipative elongation mode {11-22} is soft (semi-soft) if and only if it is marginally (or nearly marginally) stable.*

The shearing soft (semi-soft) dissipative modes $\{1,k\}$ are trivially analyzed using CE's (24b). If the shear stress components are vanished, i.e. $\sigma_{1k}^{\vec{a}} = \sigma_{1k}^{a} = 0$, the non-trivial solution of CE's (24b), the functions $\hat{e}_{1k}$ and $\hat{\omega}_{1k}$, exist only if in (24) $Det = 0$, i.e. when the condition of marginal stability (28) is satisfied. It means that the shear soft modes $\{1,k\}$ are marginally stable. On the contrary, if $Det = 0$, non-trivial solution $\{\hat{e}_{1k}, \hat{\omega}_{1k}^{r}\}$ of (24b) exists not only when $\sigma_{1k}^{\vec{a}} = \sigma_{1k}^{a} = 0$, but also when $\sigma_{1k}^{\vec{a}} \sim \sigma_{1k}^{a} \neq 0$. It means that the soft marginally stable $\{1,k\}$ modes are not unique, but along with the soft modes, non-soft but still marginally stable shear modes exist. As seen from (23), the soft shear modes are physically preferable as less energetically costly. Elementary analysis shows that the same situation holds for the semi-soft shearing modes too. This analysis is summarized as: the *soft (or semi-soft) $\{1,k\}$ dissipative shearing modes exist if and only if the marginal (or nearly marginal) stability condition (28) (or (28a)) is satisfied.*

## 6. Soft/semi-soft dissipative modes and rotational invariance

Until now, the local Cartesian axes $\vec{x}_2, x_3$ were assumed to be fixed in the plane $\{\vec{x}_2 x_3\}$ orthogonal to the director. We now consider the effect of rigid rotations of the axes $\vec{x}_2, x_3$ in the plane $\{\vec{x}_2 x_3\}$ on stress behavior and related behavior of the soft and semi-soft modes. The orthogonal matrix $\underline{\underline{q}}$ which describe these plane rotations is:

$$\underline{\underline{q}}(\alpha) = \begin{pmatrix} 1 & 0 & 0 \\ 0 & \cos\alpha & -\sin\alpha \\ 0 & \sin\alpha & \cos\alpha \end{pmatrix}.$$

Denoting the stress tensor transformed with matrix $\underline{\underline{q}}$ as $\underline{\underline{\hat{\sigma}}}' = \underline{\underline{q}}^T \cdot \underline{\underline{\sigma}} \cdot \underline{\underline{q}}$, $\underline{\underline{\hat{\sigma}}}'^T = \underline{\underline{q}}^T \cdot \underline{\underline{\sigma}}^T \cdot \underline{\underline{q}}$ and calculating the transformed stress components $\hat{\sigma}'_{11}$, $\hat{\sigma}'_{1k}$ and $\hat{\sigma}'_{k1}$ ($k = 2,3$) yields:

$$\hat{\sigma}'_{11} = \sigma_{11}, \quad \hat{\sigma}'_{12} = \sigma_{12}\cos\alpha + \sigma_{13}\sin\alpha, \quad \hat{\sigma}'_{13} = -\sigma_{12}\sin\alpha + \sigma_{13}\cos\alpha. \qquad (31a)$$

The same relations hold for $\hat{\sigma}'_{k1}$. Applying the above transformation to the system of normal stresses $\underline{\underline{\hat{\sigma}}}' = diag\{\sigma_{11}, \hat{\sigma}'_{22}, \sigma_{33}\}$ in $\{\hat{x}\}$ yields:

$$\hat{\sigma}'_{11} = \sigma_{11}; \quad \hat{\sigma}'_{22} = \sigma_{22}\cos^2\alpha + \sigma_{33}\sin^2\alpha; \quad \hat{\sigma}'_{33} = \sigma_{22}\sin^2\alpha + \sigma_{33}\cos^2\alpha. \qquad (31b)$$

Formulae (31a) show that if for particular locations of coordinate lines $\hat{x}_2, x_3$ in the plane $\{\hat{x}_2 x_3\}$ the modes $\{1,k\}$ are soft, i.e. $\{\hat{\sigma}_{1k}\} = 0$ ($k = 2, 3$), they are soft for any new Cartesian coordinates $\hat{x}'_2, x'_3$ located in the same plane. The nontrivial (i.e. related to non-zero dissipative modes) stress components are defined as *rotationally invariant* if $\forall\alpha$ they do not change their values with the rotations described by the matrix $\underline{\underline{q}}(\alpha)$. Due to (24a) it is easy to establish that the shearing stress $\hat{\sigma}_{23}$ and respective deformation mode $\{2,3\}$ cannot be rotationally invariant, but it is evident from (31a) that the *shear stresses in soft modes are rotationally invariant*. Relations (31a) for the semi-soft modes where $\hat{\sigma}_{1k}/e = O(\delta)$ ($k = 2, 3$), show that the semi-soft modes are also rotationally invariant. It means that *the set of soft (semi-soft) shearing dissipative modes is of power of continuum.* This result is almost evident for the $\{1,k\}$ soft modes because the axes $\hat{x}_2, x_3$ are located arbitrarily in the plane $\{\hat{x}_2 x_3\}$.

According to (31b) the system of stresses $\underline{\underline{\sigma}} = diag\{\sigma_{11}, \sigma_{22}, \sigma_{33}\}$ is rotationally invariant only if $\hat{\sigma}'_{22} = \sigma_{33}$. Equations (24a) show that this equality is valid only for the case of simple extension described by formula (29), being independent of softness of the stretching mode $\{11\text{-}22\}$. Therefore the *elongation mode $\{11\text{-}22\}$ is always rotationally invariant whether it is soft/semi-soft or not.*

## 7. Renormalized equations for describing soft dissipative modes

When the shearing dissipative modes $\{1,k\}$ are soft the general solution of problems for CE's (16) is complicated by the fact that the shear strains $\hat{e}_{1k}$ in the soft modes are uncertain. Nevertheless, there is another important feature of the shearing dissipative soft modes, which makes the solution of problems for CE's (16) not only unique but also much easier than in general stable case. CE's (24c) show that when the shearing modes $\{1,k\}$ are soft, the anti-symmetric components of stress tensor in $\{\hat{\underline{x}}\}$ disappear, because $\hat{\sigma}_{1k}^{a}=0$. It means that in general $\underline{\underline{\sigma}}^{a}=0$ for the viscous nematic liquids with shearing soft modes, i.e. *if the shearing modes $\{1,k\}$ are soft, the viscous stress tensor is symmetric.*

We initially use the condition $\underline{\underline{\sigma}}^{a}=0$ in general, when the asymmetric part of viscous stress could be approximately neglected. This condition is valid in the case of relatively slow motion of suspensions and/or low molecular weight liquid crystals with uniaxially symmetric particles/molecules in a viscous liquid. In the case of liquid crystals, the additional assumption for holding the condition $\underline{\underline{\sigma}}^{a}=0$ is that the anti-symmetric part of stress is possible to describe by the quasi-equilibrium (quasi-static) condition, $\underline{h}=\underline{n}(\underline{h}\cdot\underline{n})$. Then formally substituting the condition $\underline{\underline{\sigma}}^{a}=0$ in CE's (16) and (17) yields the renormalized CE's:

$$\underline{\underline{\sigma}} = \eta_0 \underline{\underline{e}} + \eta_1 (\underline{nn}\cdot\underline{\underline{e}} + \underline{\underline{e}}\cdot\underline{nn}) + 2\eta_2 \underline{nn}(\underline{nn}:\underline{\underline{e}}) \tag{32}$$

$$\overset{0}{\underline{n}} = \lambda[\underline{\underline{e}}\cdot\underline{n} - \underline{n}(\underline{nn}:\underline{\underline{e}})] \tag{33}$$

$$\underline{\underline{\omega}}^{r} = \lambda(\underline{\underline{e}}\cdot\underline{nn} - \underline{nn}\cdot\underline{\underline{e}}). \tag{34}$$

Here

$$\eta_0 = a_0, \ \eta_1 = a_1 - a_3^2/a_4, \ \eta_2 = a_2 + a_3^2/a_4, \ \lambda = a_3/a_4. \tag{35}$$

Substituting (32) into (20) results in the expressions for the renormalized dissipation:

$$D^r = \eta_0 |\underline{\underline{e}}|^2 + 2\eta_1 (\underline{nn}:\underline{\underline{e}}^2) + 2\eta_2 (\underline{nn}:\underline{\underline{e}})^2 \equiv \underline{\underline{\sigma}}_i : \underline{\underline{e}}. \tag{36}$$

Eqs. (32)-(36), resulted from the symmetric approximation of viscous stress in nematic liquids, are the Ericksen CE's whose four parameters, $\eta_0$, $\eta_1$, $\eta_2$ and $\lambda$, are presented via the five parameters in CE's (16).

The thermodynamic stability conditions for dissipation (36) is:

$$\eta_0 > 0, \quad \eta_0 + \eta_1 > 0, \quad 3/2\eta_0 + 2\eta_1 + 2\eta_2 > 0, \tag{37}$$

the parameter $\lambda$ in (32) being sign indefinite.

Substituting (35) into (37) results once again in the stability conditions (25). There might be two nontrivial dissipative marginal stability conditions, elongational,

$$3/2\eta_0 + 2\eta_1 + 2\eta_2 = 3/2a_0 + 2a_1 + 2a_2 = 0,$$

and shearing,

$$\eta_0 + \eta_1 = a_0 + a_1 - a_3^2/a_4 = 0,$$

exactly as described in the general case by Eqs.(27) and (28), respectively.

The exact, closed set of CE's (31), (32), which holds under condition of shearing marginal stability (28), is simplified to:

$$\underline{\underline{\sigma}}_i/\eta_0 = \underline{\underline{e}} - (\underline{\underline{nn}} \cdot \underline{\underline{e}} + \underline{\underline{e}} \cdot \underline{\underline{nn}}) + 1/2(1+\beta)\underline{\underline{nn}}(\underline{\underline{e}}:\underline{\underline{nn}}); \quad \overset{0}{\underline{n}} = \lambda[\underline{\underline{e}} \cdot \underline{n} - \underline{n}(\underline{\underline{nn}}:\underline{\underline{e}})]. \qquad (38)$$

Then the dissipation is represented as:

$$D^r/\eta_0 = |\underline{\underline{e}}|^2 - 2\underline{\underline{nn}}:\underline{\underline{e}}^2 + 1/2(1+2\beta)(\underline{\underline{nn}}:\underline{\underline{e}})^2 \qquad (39)$$

In eqs.(38) and (39) parameter $\beta$ is given by:

$$\beta = (3/2a_0 + 2a_1 + 2a_2)/a_0 = (3/2\eta_0 + 2\eta_1 + 2\eta_2)/\eta_0 \geq 0. \qquad (40)$$

Due to (28a) parameter $\beta$ is positive when the elongation dissipative mode {11-22} is stable. When this mode is also soft, $\beta = 0$, and the dissipative function $D^r$ in (39) is minimized. Thus the presence of both elongation and shearing dissipative modes brings the dissipation $D$ to the minimum in the parametric space $\{P\}$. In this case the CE (38) and expression for dissipation (39), scaled with the isotropic viscosity $\eta_0$, have no parameters at all, which simplifies the general flow analysis.

## 8. Examples: Simple flows of nematics with dissipative soft modes

### 8.1. Simple shearing

It is well known (see e.g. [18]) that in this flow the LC fluids usually have a strong anchoring of their molecules to the wall, with their specific orientation, which depends on the nematic type. This orientation is quite different from that induced by flow. Therefore a procedure is commonly used there to match the orientation induced by flow with the near–wall orientation described by the Frank elasticity. In the following we avoid applying this well-known matching procedure and consider only the core nematic viscous flow, mostly for given shear rate, $\dot{\gamma} = const$. It is convenient to consider the simple shearing in the common Cartesian coordinates $\{x_1, x_2, x_3\}$ where the axis $x_1$ is directed along the velocity and the axis $x_2$ along the velocity gradient.

It is easy to prove that for simple shearing, the possible steady or periodic solutions of kinematical (second) equation in (38) are two-dimensional, i.e. $n_3 = 0$. Then according to the first equation in (38), the viscous stress system is:

$$\sigma_{12}^r = 2\eta_0 \dot{\gamma}(1+\beta) n_1^2 n_2^2, \quad N_1^r = 2\eta_0 \dot{\gamma}(1+\beta) n_1 n_2 (n_1^2 - n_2^2), \quad N_2^r = 2\eta_0 \dot{\gamma} n_1 n_2 [(1+\beta) n_2^2 - 1] \tag{41}$$

Here $\sigma_{12}^r$, $N_1^r$ and $N_2^r$ are the shear stress and the first and second normal stress differences.

In the cases when $|\lambda| \neq 1$, a particular steady solution $\underline{n} = \{0, 0, 1\}$ exists which is unstable. There also exist two stable particular solutions [19], $\underline{n} = \{1, 0, 0\}$ for $\lambda = 1$ and $\underline{n} = \{0, 1, 0\}$ for $\lambda = -1$, which are not robust.

When $|\lambda| > 1$, i.e. $0 < a < 1$ (*aligning* nematics), after evolution from arbitrary initial conditions there is the steady solution:

$$n_1 = \pm\sqrt{(1+\lambda^{-1})/2}, \quad n_2 == \pm\sqrt{(1-\lambda^{-1})/2}. \quad (t \to \infty). \tag{42}$$

In this case,

$$\sigma_{12}^r = 1/2\eta_0 \dot{\gamma}(1+\beta)(1-\lambda^{-2}), \quad N_1^r = \eta_0 \dot{\gamma}(1+\beta)\sqrt{\lambda^2-1}/\lambda^2,$$

$$N_2^r = \eta_0 \dot{\gamma}\sqrt{1-\lambda^{-2}}[1/2(1+\beta)(1-\lambda^{-1})-1]. \tag{43}$$

The proportionality of stress system to $\dot{\gamma}$ for shearing flows of aligning nematics is well known for the general Ericksen equations.

When $|\lambda| < 1$ (*tumbling* nematis), steady solution is impossible. In this case the periodic solution of the second equation (38) is well known [18,20]:

$$n_1 = \left[1 + \frac{1-\lambda}{1+\lambda} \tan^2\left(\sqrt{1-\lambda^2}\dot{\gamma}t/2\right)\right]^{-1/2}, \quad n_2 = n_1\sqrt{(1-\lambda)/(1+\lambda)} \tan\left(\sqrt{1-\lambda^2}\dot{\gamma}t/2\right). \tag{44}$$

The period $T$ of nonlinear oscillations is: $T = 2\pi\left(\dot{\gamma}\sqrt{1-\lambda^2}\right)^{-1}$. When $\dot{\gamma}(t)$ is periodic, there should be a substitution: $\dot{\gamma}t \to \int \dot{\gamma}(t)dt$. In the tumbling case with $\dot{\gamma} = const$ all the components of stress system are oscillating, with their averaged values being proportional to $\dot{\gamma}$. For example, the expression for shear stress, averaged over the period of oscillations, is:

$$<\sigma_{12}^r> = 2\eta_0\dot{\gamma}(1+\beta)v(1+v)^{-2}, \quad v = \sqrt{(1-\lambda)/(1+\lambda)}. \tag{45}$$

Here the value $\eta_{ef} = 2\eta_0(1+\beta)v(1+v)^{-2}$ has a sense of effective viscosity.

It is well known that the increase in $\dot{\gamma}$ makes the effective elastic near-wall layer thinner. When the values of $\dot{\gamma}$ are high enough the non-inertial approach might be invalid. In this case, $\dot{\gamma}t \to \gamma(t, x_2)$ where

$$\gamma(t, x_2) = \int \partial_t u(x_2, t')dt', \tag{46}$$

and $u(x_2, t)$ is the space-time velocity distribution. This type of oscillations could be calculated numerically. More complicated situation occurs in inhomogeneous shearing like Poiseuille flow, when only formula (46) is generally valid. In this case the non-inertial computations with the shear stress is $\sim x_2$ seem to be not easier than the general inertial computations.

When the shear rate $\dot{\gamma}$ is very small, the actual value $\underline{n}$ of director is very close to the initial one $\underline{n}_0$. This may happen for steady shearing within the time range $\sim 1/\dot{\gamma}_0$, or for small amplitude shearing oscillations, considered for general Ericksen CE's in papers [21,22]. For instance, if $\underline{n}_0$ is directed along $x_1$ or $x_2$, it is easy to find the solution of shearing problem for homogeneous harmonic shearing oscillations, $\gamma(t) = \gamma_0 \sin(\Omega t)$, with very small amplitude $\gamma_0$ and frequency $\Omega$, as disturbances near the equilibrium state:

$$\sigma_{12}^r \sim \eta_0 \Omega \gamma_0^2 f_1(t), \quad N_1^r \sim \eta_0 \Omega \gamma_0^2 f_2(t). \tag{47}$$

Here $f_k(t)$ are periodic harmonic functions of period $\pi/\Omega$. Thus in these cases, the effects of soft (or semi-soft) dissipative nematic modes can be directly observed.

*8.2. Simple elongation*

This flow is possible only when the director is directed along the flow axis, otherwise the shearing stresses do not vanish. Easy calculations result in the following expression for the elongation stress in a steady elongation flow:

$$\sigma_{el} = \eta_0 \beta \dot{\varepsilon}, \quad D_{el} = \eta_0 \beta \dot{\varepsilon}^2. \tag{48}$$

Here $\dot{\varepsilon}$ is the elongation strain rate. When $\beta = 0$, i.e. the dissipative elongation mode {11-22} is also soft, the elongation stress and corresponding dissipation vanish. This is another case when the effects of soft (semi-soft) dissipative modes can be directly observed.

## 9. Conclusions

The dissipative soft and semi-soft modes, revealed in this paper theoretically in classical LEP theories of nematic liquids, may exist (at least for some nematic liquids) in a wide temperature range only if the parameters in these constitutive equations have the same or almost the same temperature dependencies. Unlike the elastic soft (semi-soft) deformation modes, it is impossible to directly observe their dissipative analogues in steady shearing flows, where the director orientation could not be given but is highly affected by the flow. Nevertheless it is shown that it is possible to directly observe the nematic effects of soft dissipative modes (if they exist) in steady elongation flow or in small amplitude homogeneous oscillatory shearing near the monodomain equilibrium with initial orientation of director parallel to the simple shearing axes $x_1$ or $x_2$.

It is needless to mention that the possible occurrence of the soft and semi-soft dissipative modes needs experimental verification. The authors hope that the present paper will stimulate this experimental effort.